# Statistical analysis of interplanetary shock waves measured by a Solar Wind Analyzer and a magnetometer onboard the Solar Orbiter Mission in 2023

Oleksandr Yakovlev[1*], Oleksiy Dudnik[2,1], and Anna Wawrzaszek[2]

[1]Institute of Radio Astronomy of the National Academy of Sciences of Ukraine, Kharkiv, Ukraine
Mystetstv Street, 4, Kharkiv, 61002, Ukraine

[2]Space Research Centre of the Polish Academy of Sciences, Warsaw, Poland
Bartycka Street, 18A, Warsaw, 00-716, Poland

*Corresponding author: yakovlev@rian.kharkov.ua e-mail
Contributing authors: odudnyk@cbk.waw.pl (O.D.); dudnik@rian.kharkov.ua (O.D.);
anna.wawrzaszek@cbk.waw.pl (A.W.)

Authors' ORCIDs:
Oleksandr Yakovlev: 0000-0002-4727-7678
Oleksiy Dudnik: 0000-0002-5127-5843
Anna Wawrzaszek: 0000-0001-9946-3547

Running Title: Analysis of Shock Waves Reported in 2023

## Abstract

Interplanetary (IP) shock waves are greatly interesting, as they represent significant phenomena in the near-Earth space and are direct drivers of geomagnetic and radiation storms. Moreover, various data and parameters are being explored for the identification and characterization of these waves. The spatial dimensions of shock waves vary significantly with the conditions and parameters of the propagating medium. For example, the radii of curvature of the shock wave fronts can vary by several hundred Earth radii or more in the inner heliosphere.

In this study, we improved the semi-automated identification of shock waves by analyzing the solar wind and IP magnetic field (MF) parameters. More precisely, we analyzed the data recorded by the Proton Alpha Sensor of the Solar Wind Analyzer (SWA-PAS) and magnetometer (MAG) onboard the Solar Orbiter (SolO) mission. These data were collected and analyzed during the SolO journey around the Sun in 2023 at distances of 0.29–0.95 AU.

Employing the developed algorithm, we identified over 40 IP-shock waves that occurred in 2023 using SWA-PAS and MAG. Additionally, we determined and presented a list of shock types and their basic parameters, kinetic and magnetohydrodynamic. The compression ratios, plasma beta $\beta_{us}$, angle between the shock normal and upstream MF ($\theta_{Bn}$), Mach number, and others were among these parameters. Furthermore, we investigated the statistical distributions of the $\theta_{Bn}$ and $\beta_{us}$ parameters in the upstream region. Finally, the dependence of number of the identified shock wave as a function of the distance away from the Sun was explored; the number of shocks increased gradually with the increasing heliocentric distance.

**Keywords:** coronal mass ejection, interplanetary shock wave, solar wind, space weather, Solar Orbiter.

## 1. Introduction

Coronal mass ejections (CMEs) refer to the significant release of plasma and magnetic fields (MFs, *B*) from the solar corona, accompanied by various physical effects as well as alterations of the surrounding plasma environments (Hansen 1971). Particularly, CMEs generate interplanetary (IP) shock waves and accelerate the transition of particles to higher energies (Gopalswami 2006, Yang 2024). IP shocks are common events in the circumsolar space, occurring when the CME-driven solar wind (SW) speed exceeds the sonic speed (Marcowith 2016). Shock waves can also be caused by processes occurring in the corotating interaction regions (CIRs) and streaming interaction regions (SIRs) (Gosling 1999).

Numerous studies have demonstrated IP shocks as sudden transitions between supersonic (upstream) and subsonic (downstream) flow, marked by abrupt changes in the plasma speed ($V$), density ($N$), pressure ($P$), temperature ($T$), and MF ($B$) (Trotta et al. 2024b). These changes reflect the possibility of identifying the features of IP shocks, classifying them, and studying their propagation through SW (Baumjohann 1997). Particularly, these shocks were classified as forward/reverse shocks if they were propagating antisunward/sunward in the shock frame of reference (Burlaga 1971). Moreover, fast forward (FF) shocks (Echer 2011) are characterized by increased $V$, $N$, $P$, and $B$, whereas fast reverse (FR) shocks (Kilpua 2015) are characterized by the overall decrease in all the parameters except $V$. Additionally, slow forward (SF) shocks are characterized by increased $N$, $P$, and $V$ as well as decreased $B$. Finally, slow reverse (SR) shocks are characterized by $N$ and $P$ as well as increased $B$ and $V$ (Kilpua 2015, Oliveira & Ngwira 2017, Tsurutani 2010).

Additionally, the nature of a shock wave is characterized by a set of parameters, such as, the shock normal $\mathbf{n}$ (the perpendicular direction to the shock front), the angle between $\mathbf{n}$ and the upstream $B$ ($\theta_{Bn}$), the Mach numbers ($M$; typically the Alfvenic M ($M_A$) and fast magnetosonic mode M ($M_{fms}$)), and the plasma beta ($\beta$).

Furthermore, quasiparallel and quasiperpendicular shocks represent two distinct collisionless-shock types in space plasmas, particularly from the viewpoint of SW interacting with planetary magnetospheres and IP MFs (IMFs). These shocks are classified based on the $\theta_{Bn}$ parameter. A quasiparallel shock occurs when $\theta_{Bn}$ between $B$ vector and $\mathbf{n}$ is <45°. Put differently, it occurs when $B$ is almost parallel to the shock-propagation direction. A quasiperpendicular shock occurs when $\theta_{Bn}$ between $B$ and $\mathbf{n}$ is >45°. In this case, $B$ is almost perpendicular to the shock propagation direction, assuming the shocks are planar.

Comprehensive studies based on observational data from various space missions (e.g., Wind, ACE, Ulysses, and Helios), supported by sophisticated analytical and numerical techniques, have been conducted to further elucidate the nature and effects of IP shocks as well as their dependence on solar activities (Kruparova 2013, Kilpua 2015). Numerous studies have attempted to adequately identify collisionless-shock crossings in spacecraft data. Thus, manual and automated algorithms, including machine learning (ML) approaches, have been developed for shock detection, and shock databases have been compiled for various spacecraft (Kruparova 2013, Lalti 2022, Oliveira 2023).

Particularly, Kruparova (2013) developed a method for detecting IP shock crossings using ion moments and $B$ magnitudes. Thereafter, the author verified the method using Wind (at 1 AU) and Helios (at distances of 0.29–1 AU) data. The method comprised an automated two-step detection algorithm with systematically proposed quality factors ($QF$s) and thresholds for SW parameters. Further, Cash (2014) developed an automatic IP shock detection method using eight years of ACE data to improve space weather forecasting capabilities. Notably, the heliospheric shock waves database (www.ipshocks.fi) is maintained by the University of Helsinki. This comprehensive database, which contains over 2900 fast shocks, was compiled through systematic visual inspection and ML techniques using data obtained from several missions.

The recently launched Solar Orbiter (SolO) Mission (Müller 2020), which provides high-resolution data on the plasma, $B$, and energetic particle parameters, is a crucial contributor to IP shock studies (Trotta 2023, 2024a, 2024b, Dimmock 2023). SolO enables the detection of IP shock waves at distances of 0.28–0.95 AU, with a planned maximum inclination angle of ~24°.

Following its launch in 2020, SolO has already encountered numerous heliospheric shock waves. For instance, Trotta (2023) analyzed the SolO shock observed on October 30, 2021, at 22:02:07 UT at a heliocentric distance of ~0.8 AU. The shock was a high-M IP shock with irregular "injections" of suprathermal particles generated by shock-front irregularities. The authors considered this shock an excellent avenue for studying irregular particle acceleration from a thermal plasma.

[Trotta (2024a)](#) systematically studied an observed CME-driven IP shock using the Parker Solar Probe at very small heliocentric distances (0.07 AU). CME was subsequently observed by SolO on September 6, 2022, at 10:00:51 UT at a heliocentric distance of 0.7 AU. The author revealed a very structured shock transition populated by shock-accelerated protons reaching approximately 2 MeV and exhibiting irregularities in the shock downstream, which correlated with the SW structures propagating across the shock. Furthermore, [Trotta (2024a)](#) emphasized that the local features of the IP shocks and their environments can vary significantly as they propagate through the heliosphere.

Most of the IP shock waves identified so far from the SolO data and studied systematically in the literature were observed between 2021 and 2022. Increased solar activity is anticipated as we approach the peak of solar cycle 25, and the observations in the coming years are expected to exhibit increased numbers of shocks detected by SolO. The solar cycle 25 Shocks Catalog, which was prepared in the frame of the EU Horyzon 2020 Serpentine Project (https://data.serpentine-h2020.eu) and recently published ([Trotta 2024b, Trotta 2025](#)) probably confirmed this. Conversely, identifying shocks in more complex SW states may be more challenging, although it is crucial, especially in the context of timely space weather forecasting as well as elucidating the evolution and dynamics of collisionless shocks.

Therefore, in this study, we adapted and applied [Kruparova's (2013)](#) algorithm for the detection of IP shock waves collected in the SolO data between January and December 2023 at a wide range of distances away from the Sun. Specifically, we proposed new thresholds and $QF$s for the explored plasma and $B$ parameters. Next, we systematically characterized the identified IP shocks based on a set of parameters and statistical studies as well as the dependencies of the shocks on distances, along with key information related to each shock.

Our data and methodology are presented in [Section 2](#), and the results are presented and discussed in [Section 3](#). Finally, the conclusions drawn from the study are summarized in [Section 4](#). Additionally, a full list of the shock waves and their parameters are presented in [Annexes A](#) and [B](#).

## 2. Materials and Methods

### 2.1. Solar Orbiter instruments and the data for analyzing the solar wind and interplanetary magnetic field parameters

The data recorded by the Proton Alpha Sensor of the Solar Wind Analyzer (SWA-PAS) and the magnetometer (MAG) aboard the SolO mission were used to study SW and IMF parameters.

During the exploration period (January 1 to December 31, 2023), the SolO spacecraft completed two full orbits around the Sun, and the radial distance ($R_d$) from the Sun varied from 0.29 to 0.95 AU.

Notably, SWA-PAS is an electrostatic particle analyzer [(Owen 2020)](#) that is designed to continuously measure the three-dimensional velocity-distribution function (3D-VDF) of the proton and alpha-particles as well as other parameters in the energy range of 200 eV–20 keV divided into 96 energy intervals. The azimuth view of the space was performed using 11 separate detectors and ranges of −24° to +42°. The scanning of space by the elevation angle proceeded in the ±20° range divided into nine sections. The energy adjustment and elevation-angle scanning were controlled by the electrostatic deflection system. Further, 3D-VDF of the SW protons and alpha-particle stream measurements were received every 4 or 300 s in the normal or fast mode at frequencies of 4–20 Hz, with decreasing measurement steps. The initial 3D-VDF data were processed by the SolO ground-tracking team and provided as the following SW parameters: $N$ (part/cm$^3$), velocity ($V$, km/s), $P$ (j/cm$^3$), and $T$ (eV). In this analysis, we utilized the L2 data in the SolO RTN-coordinate reference frame with a temporal resolution of 4 s.

The IMF parameters were measured by MAG ([Horbury 2020](#)) at four amplitude ranges between ±128 and ±60000 nT, with accuracy varying from 4 to 1800 nT depending on the selected range. The cadence of the $B$ measurements was determined by the selected mode; it was 1, 8, or 16 vectors per second in the normal mode and 64 or 128 vectors per second in the fast mode. For the analysis, we used the L2-level data of the IMF $B_R$, $B_T$, and $B_N$ components in the SolO RTN-coordinate reference frame.

The IMF measurements were conducted in the ±128 nT range with an accuracy of 4 nT at an acquisition rate of 8 vectors per second (measurements were performed every 0.125 s).

During the detailed examination of the L2-level measurement data, we observed that the time resolution of each deployed instrument varied throughout 2023. Therefore, to realize a unified time resolution for the data, we employed the linear-interpolation method, with reference to the SWA-PAS time scale of 4 s.

**2.2. Methodology for identifying interplanetary shock waves**

Shock wave occurrences were detected via mathematical analyses and event-signature recognition. The algorithms, which were developed, based on the results of the statistical analysis of the shock wave parameters, were used to calculate the ranges of the changes in *N*, *V*, and *T* of SW as well as the IMF magnitude to establish the criteria under which an event can be identified as a shock wave. Studies are ongoing to obtain a sufficient set of parameters for identifying shock waves and determine more advanced criteria for recording them. Moreover, this study is aimed at determining a relevant time for averaging the calculations. Shock-wave-recognition algorithms are subjected to constant modifications to optimize the shock wave detection criteria and increase the reliability of space-weather forecasting.

In the frame of this study, Kruparova's (2013) algorithm and selection criteria were adopted for the identification of IP shock waves in the SolO-recorded data. The specified algorithm was developed for the real-time recording of shock waves regardless of their type. This algorithm was adapted when developing the operation algorithm for the Selected Burst Mode 1 trigger mode of the SolO mission. The algorithm was tested via solar wind measurements over eight years (Wilson 2021). The algorithm was based on the comparison of the steps in the changes in the *N*, *V*, and IMF magnitudes with the threshold values. *QF*s of the registered event considered the entire contribution of the parameter steps in the assessment of the possibility of candidates making it into a shock wave list. *QF* must also exceed a certain threshold value. The additional criteria defined for the values of the parameter steps enabled the exclusion of weak events. The result of the algorithm (Kruparova.2013) showed that the Quality Factor (*QF1*) threshold was exceeded by 64%, *QF2* threshold was exceeded by 29% of the total number of events considered, which were subsequently considered as candidates for IP shock.

In this study, we enhanced a semi-automatic shock wave identification algorithm incorporating the simultaneous grouped changes in *N*, *V*, and IMF. The criteria for recording the shock waves based on the statistical analysis of the data derived from the SolO Mission in 2023 were refined. Moreover, the algorithm allowed for the calculation of the main parameters of the shock waves: the *B*- and *N*-compression values, *QF*s, the angle between the **n** and *B* vector, the *β* parameter, and the determination of the type of shock wave as well as its local direction.

As a first step, we inspected and selected the candidates after viewing the daily charts of the SW and IMF parameters throughout 2023.

Regarding the selected candidates, the relative changes in *N* (Δ*N*), *V* (Δ*V*), and Δ*B* at the first algorithm stage were calculated throughout the measurement interval (*Δt*), a duration of 300 s (Kruparova 2013). The calculations were conducted in a day, and the beginning of *Δt* shifted from the beginning to the end of the day. The values of the relative changes in the parameters in the measuring interval were determined, as follows:

$$\Delta N_t = \left|\frac{(N_2 - N_1)}{(N_2 + N_1)}\right| \text{ (relative change in the SW } N\text{)} \tag{1}$$

$$\Delta V_t = \left|\frac{(V_2 - V_1)}{(V_2 + V_1)}\right| \text{ (relative change in the SW } V\text{)} \tag{2}$$

$$\Delta B_t = \left|\frac{(B_2 - B_1)}{(B_2 + B_1)}\right| \text{ (relative change in the IMF intensity)}, \tag{3}$$

where the variables with indexes 1 and 2 correspond to the parameters at the beginning and end of *Δt*, respectively. The values of the relative changes in the parameters were calculated for each measurement time *t*.

Next, the obtained values of the relative changes in the selected parameters for SW and IMF were averaged throughout *Δt*, which preceded the current moment of *t*, and during *Δt* after the current moment of *t*. The averaged values of the parameters were designated as *ΔB, ΔN,* and *ΔV*.

At a further stage, the *QF*s (*QF1* and *QF2*) were determined, following Kruparova's (2013) method. The *QF* values were calculated, as follows:

$$QF1 = \frac{1}{3}\Delta B + \frac{1}{3}\Delta N + \frac{1}{3}\Delta V \qquad (4)$$

$$QF2 = \frac{1}{4}\Delta B + \frac{1}{12}\Delta N + \frac{2}{3}\Delta V. \qquad (5)$$

Another condition for identifying an event as a real shock wave was the selection of only candidates with events whose *QF1* and *QF2* exceeded a certain threshold. In turn, the threshold value of *QF*s was selected based on the analysis of the dependence of the shock wave counts on the *QF* value.

## 3. Results and Discussions

### 3.1. Selection of interplanetary shock waves from the candidates

While SW parameters were analyzed for the period considered, more than 50 candidates of the IP shock wave events were selected, for which simultaneous abrupt changes of the *ΔN, ΔV* and *ΔB* values occurred. The distributions of shock candidate events as functions of *ΔN, ΔV* and *ΔB* were calculated based on their step changes equal to 0.05.

The obtained distributions showed that most potential shocks grouped when relative changes exceeded values of 0.35 for *ΔN* (Figure 1a), 0.05 for *ΔV* (Figure 1b) and 0.30 for *ΔB* (Figure 1c). These values were chosen as thresholds for the events to be included in the list of IP shock waves.

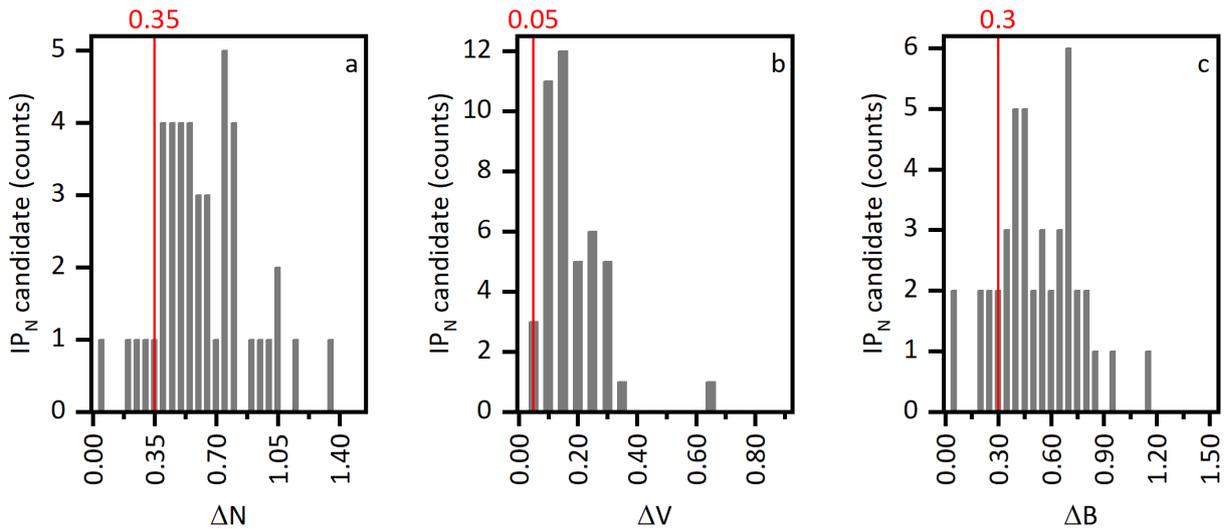

**Figure 1.** Dependence of the shock wave candidate counts (left vertical axis in all the panels) on the relative changes in the SW *N, ΔN* (a); *V, ΔV* (b); and IMF, *ΔB* (c). The red vertical lines represent the selected thresholds for identifying the shock waves: 0.35, 0.05, and 0.30 for *ΔN, ΔV,* and *ΔB,* respectively.

Regarding the final inclusion of an event as a shock wave, only such events were selected from candidate events whose *QF1* and *QF2* exceeded the threshold value. Notably, the threshold values of *QF1* and *QF2* were selected based on the results of the analysis of the dependence of the shock wave candidate count on *QF1* and *QF2* (Fig. 2). The selected threshold values were 0.25 and 0.15 for *QF1* and *QF2*, respectively, as the most counts of the shock wave candidates were recorded after exceeding these values.

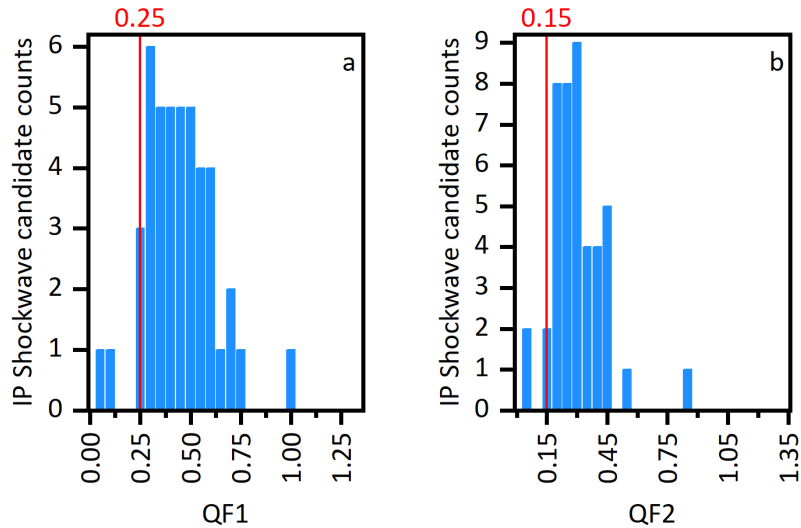

**Figure 2.** Dependence of the shock wave candidate counts (left vertical axis in all the panels) on *QF1* (a) *and QF2* (b). The red vertical lines represent the selected thresholds for identifying the shock waves.

The additional conditions for including the candidates in the list of shock waves are indicated below:
- event quality factor $Q > QF1$ or $Q > QF2$ (basic selection condition);
- simultaneous level excesses: $\Delta B > 0.1$, $\Delta N > 0.1$, and $\Delta V > 0.03$ (exclusion of events with small plasma- and IMF-parameter changes);
- $V_1 > V_2$ (exclusion of events with negative $\Delta V$).

An event was recorded as a confirmed shock wave if the event-related relative changes in all parameters simultaneously exceeded the threshold values and satisfied additional conditions. For example, Figure 3 demonstrates the visualization process for selecting doubtless shock wave candidates. Figure 3 shows a translucent red vertical strip that marks the period during which all the criteria for registering a shock wave were matched, and the red vertical dotted line shows the moment of registering the shock wave (*IP Time*) on 09.19.2023 at 02:22:59 UTC. Notably, *IP Time* was determined as the time when $\Delta B$ reached its maximum. The grey vertical translucent strip indicates the time when only some criteria were met; the event was not recorded as a shock wave.

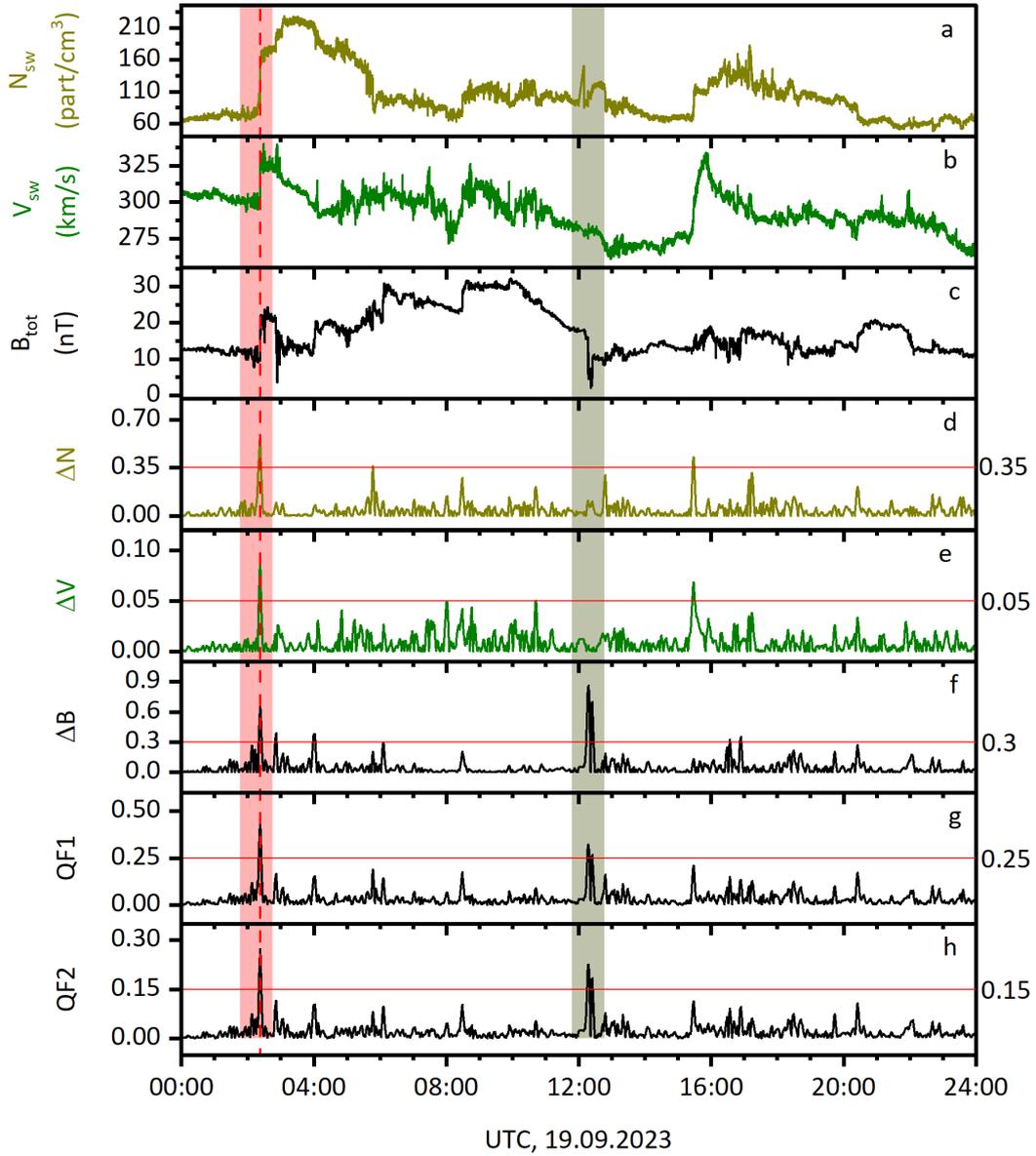

**Figure 3.** Dynamics of the SW and IMF parameters: *N* (a), *V* (b), $B_{tot}$ (c), *ΔN* (d), *ΔV* (e), and *ΔB* (f). Panels (g) and (h) display the changes in *QF1* and *QF2*. The numerical values of the parameter thresholds are displayed in the right panels and marked by red horizontal lines.

### 3.2. Basic parameters of the shock waves.

Here, 44 IP shocks (40 FF, 2 FR, 1 SF, and 1 SR) were identified using the algorithm described in Sections 2 and 3.1. For comparison, the ipshocks.helsinki.fi website documented 35 shocks for SolO in 2023 (33 FF and 2 FR), whereas the publicly accessible data through the project's Serpentine data center recorded 43 shocks (39 FF and 4 FR). The version citable through Zenodo (Trotta et al. 2024b) listed 9 shocks (all FF) for the considered period. To further characterize the registered IP shock events, a set of conventional quantitative parameters was considered. The main parameters for these events were calculated using the following widely deployed approach (e.g., Oliveira et al. 2023), and the values are listed in Tables 1 (Annex A) and 2 (Annex B).

Various studies have shown that the accuracy of the calculation of IP shock properties depends on the time periods upstream and downstream used for averaging SW parameters, and different approaches/time intervals have been used so far (Kilpua et al. 2015; Oliveira et al. 2023; Trotta et al. 2022). Specifically, Kilpua et al. (2015), using Wind data, employed an 8-minute time window for their calculations, which was based on the Helsinki database. Moreland et al. (2023), focusing on ACE measurements from 1998-2013, performed systematic studies on the effect of sampling windows on shock parameters, varying the time windows from 2 minutes to 20 minutes. Trotta et al. (2022), analyzing data from various space missions, considered different windows, with the smallest averaging windows being about 2 minutes upstream and downstream and the largest windows being about 10 minutes. They suggested that the extent and location of the best upstream and downstream analysis intervals vary depending on the shock.

In this work, various time windows determining the upstream (non-shocked) and downstream (shocked) periods (ranging from 5 to 10 minutes) were initially studied for their shock properties. Finally, the fixed and equal upstream and downstream intervals of 300 s (5 minutes, corresponding to 75 periods of measuring the SW parameters) were applied. These intervals were selected taking into account the shock durations, the high data resolution of the considered SolO data, the correctness of the MHD description, and, finally, their length being sufficient to average out turbulence and wave activity for as many shocks as possible.

The $B$ and $N$ ratios ($r_B$ and $r_N$) were defined as the downstream mean value of the time window divided by the upstream mean value of the same time window. These ratios are expressed, as follows:

$$r_B = \frac{\langle B_{ds} \rangle}{\langle B_{us} \rangle}, \tag{6}$$

and

$$r_N = \frac{\langle N_{ds} \rangle}{\langle N_{us} \rangle}. \tag{7}$$

The upstream $\beta$ ($\beta_{us}$) parameter (the ratio of the plasma to magnetic $P$) was computed to characterize the upstream zone, as follows:

$$\beta_{us} = 2\mu_0 N_{us} k_B T / B_{us}^2, \tag{8}$$

where $\mu_0$ is the magnetic vacuum permeability and $k_B$ is the Boltzmann constant (e.g., Burgess 1993, Alexandrova et al. 2007). The Alfvenic $V$ ($V_A$) represents another parameter in Tables 1–2 that depends on the $B$ and plasma variables; this parameter was considered in the upstream region and expressed, as follows:

$$V_A = B_{us}/\sqrt{\mu_0 N_{us} m_p}), \tag{9}$$

where $m_p$ is the proton mass.

The normal vector to the shock wave front ($\boldsymbol{n}$), calculated using the MX3-mode method (Abraham-Shrauner 1976, Schwartz 1998), was subsequently applied to determine $\theta_{Bn}$, i.e., the angle between the upstream magnetic field $B_{us}$ and the $\boldsymbol{n}$ vectors. Additionally, $\theta_{Bn}$ controls the behavior of the particle event during the shock (Kilpua et al. 2015) and was computed, as follows:

$$\theta_{Bn} = \frac{180°}{\pi} \arccos(\frac{\boldsymbol{n} * \boldsymbol{B_{us}}}{|\boldsymbol{B_{us}}|}). \tag{10}$$

In the next step, the shock speed, $V_{sh}$, was evaluated thus:

$$V_{sh} = \left| \frac{N_{ds}\boldsymbol{V_{ds}} - N_{us}\boldsymbol{V_{us}}}{N_{ds} - N_{us}} \boldsymbol{n} \right|. \tag{11}$$

The sound velocity in the IP space $C_s$ is expressed thus:

$$C_s^2 = \gamma(T_e + T_p)/m_p, \tag{12}$$

where the adiabatic index ($\gamma$) = 5/3 and electron temperature ($T_e$) were calculated, following Chart's (2011) method. More specifically, the following relation for $T_e$ was used for varied SolO locations:

$$T_e = 146277\, R_d^{(-0.664)}, \qquad (13)$$

where coefficient 146277 represents $T_e$ (in K) at 1 AU, and $R_d$ denotes the radial distance from the Sun (in AU). The fast magnetosonic and Afvénic Ms ($M_{fms}$ and $M_A$, respectively), as crucial characteristics of the intensity of the identified shocks, were also calculated, following Schwartz (1998):

$$M_A = \frac{|\mathbf{V}_{us}\,\mathbf{n} \pm V_{sh}|}{V_A} \qquad (14)$$

and

$$M_{fms} = \frac{|\mathbf{V}_{us}\,\mathbf{n}|}{V_{fms}}, \qquad (15)$$

where $V_{fms}$ is defined as the positive value of the expression:

$$V_{fms} = \frac{1}{2}\sqrt{V_A^2 + C_s^2 \pm \sqrt{(V_A^2 + C_s^2)^2 - 4 V_A^2 C_s^2 \cos^2\theta_{Bn}}}. \qquad (16)$$

## 3.3. Prompt analysis of the heliocentric characteristics of the selected parameters of the shock waves

Based on the results obtained thus far, we evaluated the dependences of the *N*- and *B*-compression parameters as well as *QF*s as functions of the distance $R_d$ between the SolO and Sun, and the results are shown in Figure 4.

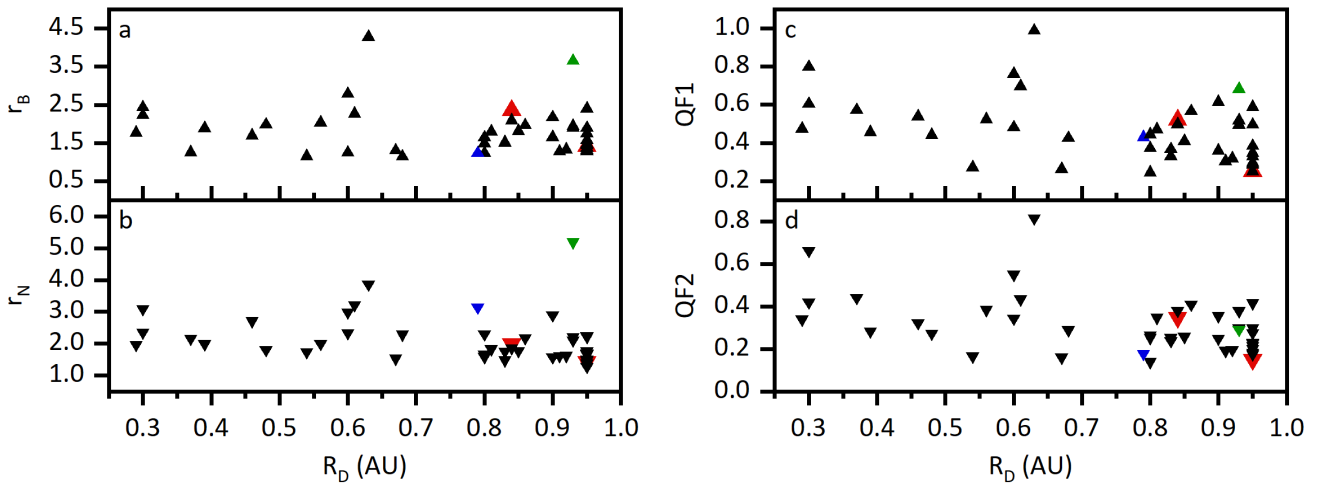

**Figure 4.** Dependences of $r_B$ (a), $r_N$ (b), *QF1* (c), and *QF2* (d) as functions of the distance between the SolO and Sun for each shock wave presented in Table 1. The black points denote FF, red; FR, blue; SF, green; and SR, IP shocks.

The obtained dependencies confirmed that the main parameters of the IP shock reached their maximum values within the measurement period at a distance of ~0.63 AU between the spacecraft and the Sun, and this is significantly associated with one specific powerful event in the IP space; thus, it was not a statistically obtained average result. The distributions of the shock wave count determined as a function of $\theta_{Bn}$ (Fig. 5a) and depended on $\beta_{us}$ (Fig. 5b).

Figure 5a shows the largest counts of the undoubtedly shock waves were recorded at the SolO location, and it was quasiperpendicular, with an angle ($\theta_{Bn}$) of ~50°.

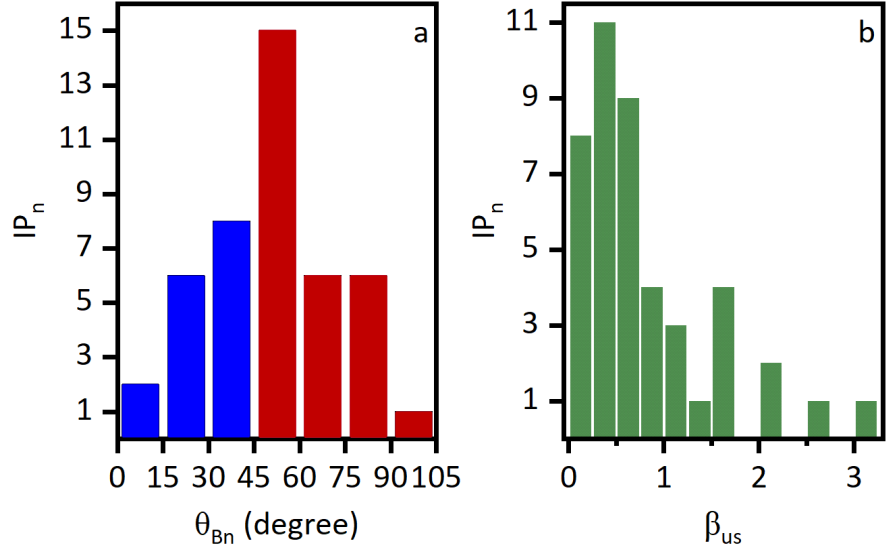

**Figure 5.** Distribution of the $IP_n$ shock counts as functions $\theta_{Bn}$ (a) and $\beta_{us}$ (b). Blue bins denote $IP_n$ with $\theta_{Bn}$ <45°, while red bins correspond to $\theta_{Bn}$ >45°.

Figure 5a and values of $\theta_{Bn}$ given in Table 2 suggest that the most shock waves recorded at the SolO location in 2023 can be considered as quasi-perpendicular ($\theta_{Bn}$>45°, red bins in Fig. 5a), with the dominant number of cases around ~50°.

The histogram in Fig. 5b indicates that the majority of the registered shock waves are characterized by $\beta_{us}$ values less than 1, highlighting the dominant role of magnetic field in dynamics compared to thermal motion in these cases. The shock waves with $\beta_{us}$ >1 were also observed in 8 out of 44 cases, but the maximum values of $\beta_{us}$ did not exceed 3.25.

**3.4. Distribution of the shock waves along the Solar Orbiter trajectory**

As SolO is located at various distances from the Sun at each moment, elucidating the dependence of the number and intensity of the identified shock waves on the heliocentric distance was crucial. To visualize each shock wave along the trajectory of the mission, *QF1* and *QF2* were deployed as indicators of the intensity of the event, as they simultaneously contain information about *ΔN*, *ΔV*, and *ΔB*; the elucidated dependencies are shown in Figure 6.

To explore the sources of the shock waves in the inner heliosphere, the fast coronal mass ejections (CMEs), or solar wind stream interaction regions (SIRs) must be considered. To study this relationship, we compared the time moments of solar flare occurrence (STIX Data Center)[1] with the shock waves identified during the SolO travel time, and the resulting dependence is displayed in Figure 6c. The red triangles indicate the registered X-class solar flares, and the M- and C-class solar flares are shown in green and blue, respectively. Here, we did not consider B-class solar flares because of their lower efficiency to produce various manifestations in the interplanetary space, including CMEs, as compared with M- and X- classes of solar flares.

---
[1] https://datacenter.stix.i4ds.net/view/ancillary

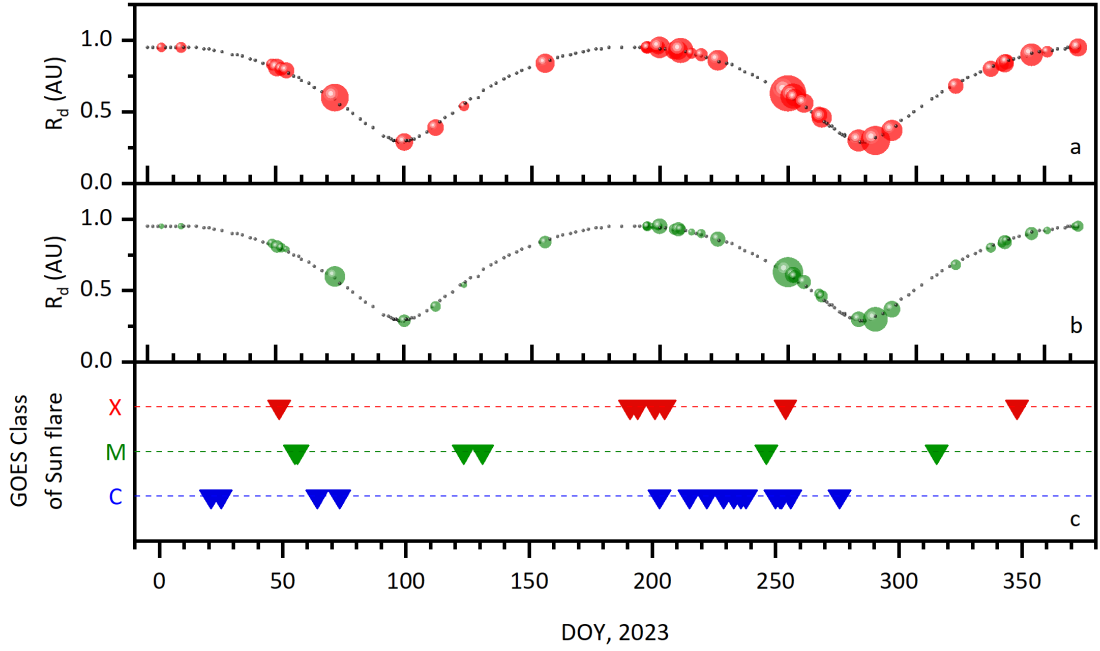

**Figure 6.** Distributions of the shock waves recorded by the SWA-PAS and MAG instruments in 2023 along the SolO trajectory compared with the X-, M-, and C-class solar flares obtained from the STIX Data Center. The dark dots in panels *a* and *b* indicate the daily average (UTC ≈ 12:00) spacecraft location according to the Solar-MACH catalog[2], with a time step of 1–5 days. The top panel shows the moments of registering the shock waves for *QF1* in the form of spheres of different diameters; the middle panel displays the same information as Panel *a* but for *QF2*. The diameters of the spheres were proportional to the *QF1* and *QF2* values listed in Table 1. Panel c shows the time points at which different-class X-ray flares were recorded.

The CMEs, in their turn, may produce shock waves in the interplanetary space. In summary, 16, 6, and 7 C-, M-, and X-class flares, respectively, were recorded in 2023 (STIX Data Center, DONKI Database[3]). Figure 6c shows that the following solar flares were recorded in a $R_d$ of 0.55–0.65 AU: 5, 1, and 1 C-, M2.1-, and X4-class flares, respectively. We assume that these solar flares could induce local amplification in $IP_n$ and $N_{IP\_Rd}$ in the $R_d$ segment ($R_d$ = 0.55…0.65 AU; Fig. 7a).

The determination of the dependence of the number of shock waves on the heliocentric distance requires the consideration of the time spent by the spacecraft on each trajectory segment to normalize the data.

We examined the distribution of the number of identified shock waves based on the distance between the Sun and the spacecraft throughout 2023. During this period, SolO completed two orbits around the Sun. Notably, the minimum and maximum $R_d$ were 0.29 and 0.95 AU, respectively. To analyze the events, the full $R_d$ range was divided into 10 equal segments (length = 0.067 AU). The moment at which the spacecraft entered the segment and the time spent on its passage were determined using the Solar Orbiter Archive data[4], and the results are shown in Figure 7.

---

[2] https://solar-mach.github.io
[3] https://ccmc.gsfc.nasa.gov/tools/DONKI/
[4] https://soar.esac.esa.int/soar/

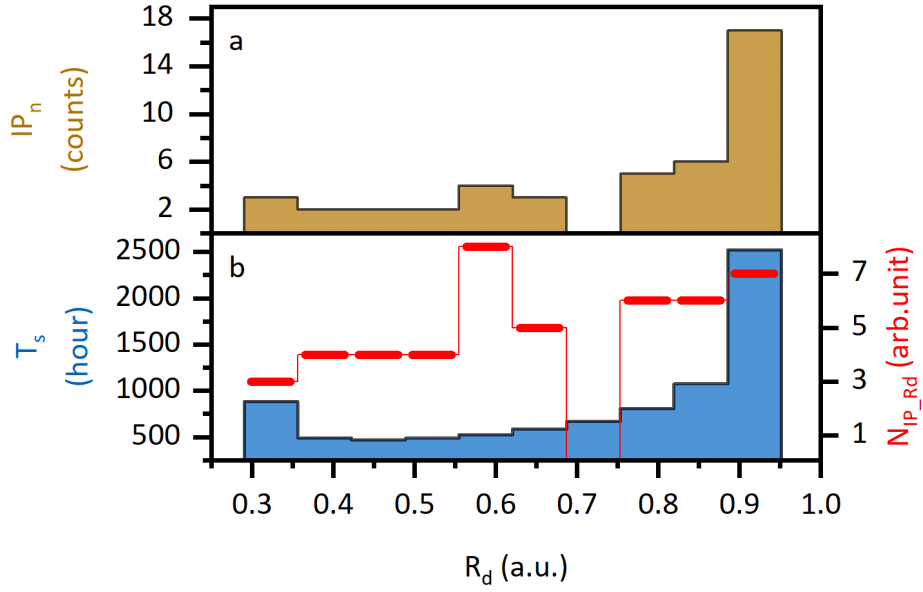

**Figure 7.** a) Dependence of the number of shock waves ($IP_n$) as a function of $R_d$; b) Dependence of the time spent on passing a particular segment (length = 0.067 AU; $T_s$) as a function of $R_d$; $N_{IP\_Rd}$ is the ratio of $IP_n$ on each segment ($IP_s$) to $T_s$ as a function of $R_d$.

The dependence of $N_{IP\_Rd}$ on the time the spacecraft passes this segment as a function of $R_d$ ($N_{IP\_Rd}$) is derived, as follows:

$$N_{IP\_Rd} = \frac{IPn}{Ts}. \tag{17}$$

Figure 7a shows that the total (during two orbits of the spacecraft around the Sun) $IP_n$ increased with the increasing $R_d$. In this case, the total time spent by the spacecraft before passing successive segments $R_d$ with an increasing $R_d$ also increased (Fig. 7b). One of the reasons for the change in $T_s$ might be the decrease in $V$ of the spacecraft as it moves away from the Sun as well as an increase in $V$ as $R_d$ decrease (Solar Orbiter Archive; Solar-MACH catalogue).

The nature of the change in the $N_{IP\_Rd}$ parameter indicates that $IP_n$ tended to increase as $R_d$ increased from the minimum to the maximum. One of the factors accounting for this trend at small $R_d$ values is the decrease in the area of spatial coverage of the IP space in the direction of the Sun by the SWA-PAS and MAG instruments as well as the decrease in $IP_n$ driven by CME.

As the spacecraft moved further away from the Sun with the $R_d$ value approaching 1 AU, the spatial viewing area in the direction of the Sun increased gradually, thus increasing the number of shock waves. The formation of CIR/SIR regions proceeded at distances exceeding 1 AU from the Sun. Shock waves will occur if the difference in the high and low SW velocities in the stream interface region of the interacting CIRs/SIRs reached the local-magnetosonic velocity and exceeds it. Such events typically occurred at distances of ~1.5 AU and were studied by the Ulysses mission (Heber 1999).

Additionally, the analysis depicted in Figure 7 revealed another effect: a slightly larger number of IP shocks occurred during the second half of 2023. Notably, Dimmock et al. (2023) deployed an automated search method adapted from the IP shock database maintained by the University of Helsinki (Kilpua et al. 2015) to perform IP shock detection from the beginning of the SolO mission (February 2020) until August 31, 2022, identifying a total of 47 shocks. Comparing this result with the $IP_n$ identified in 2023 reveals that the fraction of FF shocks increased during the ascending phase of solar cycle 25. This effect correlates with those reported by Kilpua *et al.* (2015) and Oliveira *et al.* (2023) for other solar cycles or different missions.

4.     **Conclusions**

In this study, we enhanced a semi-automatic algorithm for identifying IP shock waves for a small statistical sample during the ascending phase of the 25$^{th}$ solar-activity cycle. The algorithm enabled the selection of events that were identified as IP-shock waves. Employing the algorithm, the moments of *in situ* shock wave registrations were determined by analyzing $\Delta N$, $\Delta V$, and $\Delta B$ of SW, as well as the IMF intensity and *QF1* and *QF2*. When developing the algorithm, the threshold values of $\Delta N$, $\Delta V$, and $\Delta B$ were determined, and they supported the identification of shocks in the following years of the SolO mission.

Employing the proposed algorithm, 44 various IP shock waves that occurred in the inner heliosphere in 2023 were identified. Most of them, i.e., 40, were identified as FF-type shock waves, two were identified as FR-type shock waves, and one each was identified as SF- and SR-type shock waves. Further, we calculated the typical kinetic and magnetohydrodynamic characteristics of each identified shock wave. The dependences of $r_N$ and $r_B$ on the heliocentric distance indicated that the typical values of $r_N$ for the FF-type shocks changed from 1.5 to 3.9, whereas the $r_B$ values for the same group of shock waves varied from 1.2 to 4.3.

Furthermore, $\theta_{Bn}$ in the upstream region revealed that $\theta_{Bn}$ for most shock waves identified in 2023 ranged from 15° to 90°, with a maximum of 45°–60°. Namely, the FF-type shock waves were quasiperpendicular in most cases. The distribution of the $\beta_{us}$ parameter demonstrated that $\beta_{us} < 1$ for most shock waves, with values exceeding 1 in only 8 cases. This suggests a dominant role of magnetic field in the dynamics compared to thermal motion for IP shocks recorded at the SolO location in 2023.The nature of the change in the $N_{IP\_Rd}$ parameter indicated that the number of shock waves increased with the increasing distance ($R_d$) from the minimum to the maximum. The reasons for this tendency at small $R_d$ are the limited area for the spatial observation of the IP space by the SWA-PAS and MAG instruments and the failure to differentiate $V$ of the high and low SW $V$ of the local-magnetosonic wave.

**Acknowledgments**

This study was supported by the "Long-term program of support of the Ukrainian research teams at the Polish Academy of Sciences conducted in collaboration with the U.S. National Academy of Sciences and financially supported by external partners" (Agreement No. PAN.BFB.S.BWZ.363.022.2023).

The authors are grateful to the scientific and technical group of the SWA and MAG teams of the SolO Mission. The authors acknowledge Prof. C.J. Owen (MSSL-UCL, the University College, London, U.K.) and Prof. T. Horbury (Blackett Laboratory, Imperial College, London, U.K.) for generating the data from SWA and MAG.

The authors appreciate "Enago" (www.enago.com) for the English language review.

**Data availability**

The data employed for this study are available in the following open sources:
1. The data from STIX Data Center are available at https://datacenter.stix.i4ds.net/.
2. The data from Solar-MACH catalog are available at https://solar-mach.github.io.
3. The data from Space-Weather Database of Notifications, Knowledge, Information DONKI Database are available at https://ccmc.gsfc.nasa.gov/tools/DONKI/.
4. The data from Solar Orbiter Archive are available at https://soar.esac.esa.int/soar/#home.

**References**

Abraham-Shrauner, B, Yun, SH. 1976. Interplanetary shocks seen by Ames plasma probe on Pioneer 6 and 7. *J. Geophys. Res.*, **81:** 2097-2102. https://doi.org/10.1029/JA081i013p02097.

Alexandrova, O, Carbone, V, Veltri, P, Sorriso-Valvo, L. 2007. Solar wind cluster observations: turbulent spectrum and role of hall effect. *Planet. Space Sci.*, **55**: 2224-2227. https://doi.org/10.1016/j.pss.2007.05.022.

Baumjohann, W, Treumann, RA. 1997. Basic Space Plasma Physics, Imperial College Press, London, UK. https://doi.org/10.1142/p020.

Burgess, D. 1993. Collisionless shocks, in Introduction to Space Physics, edited by MG Kivelson and CT Russell, Cambridge University Press, Cambridge, UK.

Burlaga, LF. 1971. Hydromagnetic waves and discontinuities in the solar wind. *Space Sci. Rev.*, **12(5)**, 600-657. https://doi.org/10.1007/BF00173345.

Cash, MD, Wrobel, JS, Cosentino, KC, Reinard, AA. 2014. Characterizing interplanetary shocks for development and optimization of an automated solar wind shock detection algorithm. *J. Geophys. Res. Space Phys.*, **119**: 4210-4222. https://doi.org/10.1002/2014JA019800.

Chat, GL, Issautier, K, Meyer-Vernet, N, Hoang, S. 2011. Large-scale variation of solar wind electron properties from quasi-thermal noise spectroscopy: ulysses measurements. *Solar Phys,* **271**: 141-148. https://doi.org/10.1007/s11207-012-9967-y.

Dimmock, AP, Gedalin, M, Lalti, A, Trotta, D, Khotyaintsev, YV, Graham, DB, Johlander, A, Vainio, R, Blanco-Cano, X, Kajdič, P, Owen, CJ, Wimmer-Schweingruber, RF. 2023. Backstreaming ions at a high Mach number interplanetary shock. *Astron. Astrophys.*, **679**: A106. https://doi.org/10.1051/0004-6361/202347006.

Echer, E, Tsurutani, BT, Guarnieri, FL, Kozyra, JU. 2011. Interplanetary fast forward shocks and their geomagnetic effects: CAWSES events. *J. Atmos. Sol.-Terr. Phys.*, **73:** 1330-1338. https://doi.org/10.1016/j.jastp.2010.09.020.

Gopalswamy, N. 2006. Properties of interplanetary coronal mass ejections. 2006. *Space Sci. Rev.,* **124**: 145-168. https://doi.org/10.1007/s11214-006-9102-1.

Gosling, JT, Pizzo, VJ. 1999. Formation and evolution of corotating interaction regions and their three dimensional structure. *Space Sci. Rev.,* **89:** 21-52. https://doi.org/10.1023/A:1005291711900.

Hansen RT, Garcia CJ, Grognard, RJM, Sheridan, KV. 1971. A coronal disturbance observed simultaneously with a white-light corona-meter and the 80 MHz Culgoora radioheliograph. *Publ. Astron. Soc. Aust.,* **2:** 57-60. https://doi.org/10.1017/S1323358000012856.

Heber, B, Sanderson, TR, Zhang, M. 1999. Corotating interaction regions. *Adv. Space Res*. **23:** 567-579. https://doi.org/10.1016/S0273-1177(99)80013-1.

Horbury, TS, O'Brien, H, Carrasco Blazquez, I, Bendyk, M, Brown, P, Hudson, R, Evans, V, Oddy, TM. 2020. The Solar Orbiter magnetometer, *Astron. Astrophys.,* **642**: A9. https://doi.org/10.1051/0004-6361/201937257.

Kilpua, EKJ, Lumme, E, Andreeova, K, Isavnin, A, Koskinen, HEJ. 2015. Properties and drivers of fast interplanetary shocks near the orbit of the Earth (1995–2013). *J. Geophys. Res. Space Phys.*, **120**: 4112-4125. https://doi.org/10.1002/2015JA021138.

Kruparova, O, Maksimovic, M, Šafránková, J, Němeček, Z, Santolík, O, Krupar, V. 2013. Automated interplanetary shock detection and its application to Wind observations. *J. Geophys. Res. Space Phys.*, **118**: 4793-4803. https://doi.org/10.1002/jgra.50468.

Lalti, A, Khotyaintsev, Yu.V, Dimmock, AP, Johlander, A, Graham, DB, Olshevsky, V/ 2022. A database of MMS bow shock crossings compiled using machine learning. *J. Geophys. Res. Space Phys.*, **127**. https://doi.org/10.1029/2022JA030454.

Marcowith, A, Bret, A, Bykov, A, Dieckman, ME, Drury, LO'C, Lemb`ege, B, Lemoine, M, Morlino, G, Murphy, G, Pelletier, G, Plotnikov, I, Reville, B, Riquelme, M, Sironi, L, Stockem Novo, A. 2016. The microphysics of collisionless shock waves. *Rept. Prog. Phys.*, **79**: 046901. https://doi.org/10.1088/0034-4885/79/4/046901.

Moreland, K., Dayeh, M.A., Li, G., Farahat, A., Ebert, R.W., Desai, M.I. 2023. Variability of interplanetary shock and associated energetic particle properties as a function of the time window around the shock. *ApJ,* **956:** 107. https://iopscience.iop.org/article/10.3847/1538-4357/acec6c.

Muller, D, Cyr, OCSt, Zouganelis, I, Gilbert, HR, Marsden, R, Nieves-Chinchilla, T, et al. 2020. The solar orbiter mission science overview, *Astron. Astrophys.*, **642**: A1. https://doi.org/10.1051/0004-6361/202038467.

Oliveira, DM. 2023. Interplanetary shock data base, *Front. Astron. Space Sci.* **10**: 1240323. https://doi.org/10.3389/fspas.2023.1240323.

Oliveira, DM, Ngwira, CM. 2017. Geomagnetically induced currents: principles. *Braz. J. Phys.*, **47**: 552-560. https://doi.org/10.1007/s13538-017-0523-y.

Owen, CJ, Bruno, R, Livi, S, Louarn, P, Al Janabi, et al. 2020. The solar orbiter solar wind analyser (swa) suite. *Astron. Astrophys.*, **642**: A16. https://doi.org/10.1051/0004-6361/201937259.

Schwartz, SJ. 1998. Shock and Discontinuity normals, mach numbers, and related parameters, in *Analysis Methods for Multi-Spacecraft Data, Electronic Edition, July 2000,* edited by Paschmann, G and Daly, W, International Space Science Institute.

Trotta, D, Hietala, H, Dresing, N, Horbury, T, Kartavykh, Y, Gieseler, J, Jebaraj, IC, Gómez-Herrero, R, Espinosa Lara, F, Vainio, R. 2024b. Solar orbiter cycle 25 interplanetary shock list [Data set]. Zenodo. https://doi.org/10.5281/zenodo.12518015.

Trotta, D., Vuorinen, L., Hietala, H., Horbury, T., Dresing, N., Gieseler, J., Kouloumvakos, A., Price, D., Valentini, F., Kilpua. E., Vainio, R. 2022. Single-spacecraft techniques for shock parameters estimation: A systematic approach. *Front. Astron. Space Sci.,* **9:**1005672. https://doi.org/10.3389/fspas.2022.1005672.

Trotta, D, Horbury, TS, Lario, D, Vainio, R, Dresing, N, Dimmock, A, Giacalone, J, Hietala, H, Wimmer-Schweingruber, RF, Berger, L, Yang, L. 2023. Irregular proton injection to high energies at interplanetary shocks. *Astrophys. J. Lett.*, **957**: L13 (5pp). https://doi.org/10.3847/2041-8213/ad03f6.

Trotta, D, Larosa, A, Nicolaou, G, Horbury, TS, Matteini, L, Hietala, H, Blanco-Cano, X, Franci, L, Chen, CH K, Zhao, L. 2024a. Properties of an interplanetary shock observed at 0.07 and 0.7 au by parker solar probe and solar orbiter. *Astrophys. J.*, **96**: 2, 147. https://doi.org/10.3847/1538-4357/ad187d.

Trotta, D, et al 2025. An Overview of Solar Orbiter Observations of Interplanetary Shocks in Solar Cycle 25. *The Astrophysical Journal Supplement Series*, **277**, 2. https://doi.org/10.3847/1538-4365/ada4a7.

Tsurutani, BT, Lakhina, GS, Verkhoglyadova, OP, Gonzalez, WD, Echer, E, Guarnieri, FL. 2010. A review of interplanetary discontinuities and their geomagnetic effects. *J. Atmos. Sol.-Terr. Phys.,* **73**: 5-19. https://doi.org/10.1016/j.jastp.2010.04.001.

Wilson, LB, Brosius, AL, Gopalswamy, N, Nieves-Chinchilla, T, Szabo, A, et al. 2021. A quarter century of wind spacecraft discoveries. *Rev. Geophys.,* **59**: e2020RG000714**.** https://doi.org/10.1029/2020RG000714.

Yang, L, Heidrich-Meisner, V, Wang, W, Wimmer-Schweingruber, RF, Wang, L, Kollhoff, A, Berger, L, Pacheco, D, Xu, Z, Rodríguez-Pacheco, J, Ho, GC. 2024. Dynamic acceleration of energetic protons by an interplanetary collisionless shock. *A&A*, **686**: A132. https://doi.org/10.1051/0004-6361/202348723.

**Annex A**
Table 1

| 1 | 2 | 3 | 4 | 5 | 6 | 7 | 8 | 9 | 10 | 11 |
|---|---|---|---|---|---|---|---|---|---|---|
| № | Date [dd/mm] | IP Time [UTC] | Dist. [AU] | Angle S–E [°] | $r_B$ | $r_N$ | $QF1$ | $QF2$ | IP Type | $\beta_{us}$ |
| 1 | 05/01 | 09:44:07 | 0.95 | −22.4 | 1.42 | 1.44 | 0.25 | 0.15 | FR | 1.31 |
| 2 | 13/01 | 01:37:53 | 0.95 | −24.7 | 1.32 | 1.70 | 0.30 | 0.18 | FF | 0.15 |
| 3 | 17/02 | 14:03:34 | 0.83 | −32.6 | 1.52 | 1.49 | 0.33 | 0.25 | FF | 0.47 |
| 4 | 19/02 | 09:58:46 | 0.81 | −32.7 | 1.80 | 1.83 | 0.47 | 0.35 | FF | 0.28 |
| 5 | 20/02 | 22:48:51 | 0.80 | −32.7 | 1.24 | 1.59 | 0.25 | 0.14 | FF | 0.22 |
| 6 | 21/02 | 03:35:54 | 0.80 | −32.7 | 1.64 | 1.66 | 0.37 | 0.25 | FF | 0.28 |
| 7 | 23/02 | 03:49:07 | 0.79 | −32.7 | 1.25 | 3.18 | 0.44 | 0.18 | SF | 0.14 |
| 8 | 14/03 | 01:08:26 | 0.60 | −25.8 | 2.79 | 2.98 | 0.76 | 0.55 | FF | 0.84 |
| 9 | 10/04 | 04:33:38 | 0.29 | 62.9 | 1.77 | 1.96 | 0.48 | 0.34 | FF | 0.73 |
| 10 | 22/04 | 08:02:08 | 0.39 | −232.9 | 1.89 | 2.00 | 0.46 | 0.28 | FF | 0.48 |
| 11 | 03/05 | 10:25:17 | 0.54 | 151.4 | 1.16 | 1.73 | 0.27 | 0.17 | FF | 0.02 |
| 12 | 04/06 | 03:07:10 | 0.84 | 165.0 | 2.36 | 2.01 | 0.52 | 0.35 | FR | 2.66 |
| 13 | 13/07 | 17:32:56 | 0.95 | −201.8 | 1.37 | 1.68 | 0.33 | 0.23 | FF | 0.29 |
| 14 | 13/07 | 20:38:04 | 0.95 | −201.8 | 1.47 | 1.38 | 0.29 | 0.23 | FF | 0.13 |
| 15 | 14/07 | 00:30:17 | 0.95 | −201.9 | 1.44 | 1.51 | 0.35 | 0.28 | FF | 0.28 |
| 16 | 16/07 | 03:22:25 | 0.95 | −202.4 | 1.57 | 1.28 | 0.25 | 0.18 | FF | 0.19 |
| 17 | 18/07 | 06:09:41 | 0.95 | −202.8 | 1.75 | 1.77 | 0.39 | 0.23 | FF | 0.95 |
| 18 | 18/07 | 08:58:33 | 0.95 | −202.9 | 1.44 | 1.47 | 0.29 | 0.20 | FF | 0.11 |
| 19 | 18/07 | 17:06:49 | 0.95 | −202.9 | 2.40 | 2.24 | 0.59 | 0.42 | FF | 1.69 |
| 20 | 24/07 | 11:59:58 | 0.93 | 155.8 | 1.89 | 2.11 | 0.50 | 0.30 | FF | 1.11 |
| 21 | 26/07 | 01:22:46 | 0.93 | 155.5 | 1.94 | 2.21 | 0.52 | 0.38 | FF | 2.20 |
| 22 | 27/07 | 00:38:42 | 0.93 | 155.3 | 3.65 | 5.19 | 0.68 | 0.29 | SR | 1.11 |
| 23 | 31/07 | 06:39:03 | 0.91 | 154.1 | 1.28 | 1.61 | 0.30 | 0.19 | FF | 0.57 |
| 24 | 04/08 | 01:29:15 | 0.90 | 153.9 | 1.66 | 1.58 | 0.36 | 0.25 | FF | 0.60 |
| 25 | 10/08 | 11:48:08 | 0.86 | −206.9 | 1.97 | 2.18 | 0.57 | 0.41 | FF | 0.57 |
| 26 | 03/09 | 07:12:56 | 0.67 | 156.3 | 1.31 | 1.54 | 0.26 | 0.16 | FF | 0.70 |
| 27 | 06/09 | 20:23:33 | 0.63 | −201.7 | 4.28 | 3.87 | 0.99 | 0.82 | FF | 3.04 |
| 28 | 08/09 | 19:32:13 | 0.61 | −200.2 | 2.27 | 3.22 | 0.70 | 0.43 | FF | 2.12 |
| 29 | 09/09 | 10:41:49 | 0.60 | −199.7 | 1.25 | 2.33 | 0.48 | 0.34 | FF | 0.24 |
| 30 | 13/09 | 03:13:38 | 0.56 | −196.1 | 2.03 | 2.00 | 0.53 | 0.38 | FF | 0.71 |
| 31 | 19/09 | 02:22:59 | 0.48 | 172.9 | 1.98 | 1.81 | 0.44 | 0.27 | FF | 1.62 |
| 32 | 20/09 | 00:47:19 | 0.46 | 174.8 | 1.70 | 2.71 | 0.54 | 0.32 | FF | 0.33 |
| 33 | 04/10 | 11:28:17 | 0.30 | −126.5 | 2.24 | 2.35 | 0.61 | 0.42 | FF | 0.49 |
| 34 | 10/10 | 22:32:22 | 0.30 | −84.6 | 2.44 | 3.09 | 0.80 | 0.66 | FF | 0.96 |
| 35 | 17/10 | 10:23:59 | 0.37 | −52.3 | 1.26 | 2.16 | 0.57 | 0.44 | FF | 0.65 |
| 36 | 11/11 | 09:14:07 | 0.68 | −12.7 | 1.15 | 2.29 | 0.43 | 0.29 | FF | 1.66 |
| 37 | 25/11 | 00:30:53 | 0.80 | −10.0 | 1.48 | 2.30 | 0.45 | 0.26 | FF | 0.57 |
| 38 | 29/11 | 07:51:09 | 0.83 | −10.0 | 1.51 | 1.76 | 0.37 | 0.24 | FF | 0.46 |
| 39 | 30/11 | 10:47:26 | 0.84 | −10.1 | 2.09 | 1.86 | 0.50 | 0.38 | FF | 0.42 |
| 40 | 01/12 | 02:26:40 | 0.85 | −10.2 | 1.82 | 1.77 | 0.41 | 0.26 | FF | 0.57 |
| 41 | 10/12 | 21:14:30 | 0.90 | −11.5 | 2.17 | 2.90 | 0.62 | 0.36 | FF | 1.63 |
| 42 | 17/12 | 02:08:54 | 0.92 | −12.9 | 1.33 | 1.62 | 0.32 | 0.20 | FF | 0.46 |
| 43 | 27/12 | 22:22:22 | 0.95 | −15.7 | 1.29 | 1.66 | 0.33 | 0.22 | FF | 0.94 |
| 44 | 29/12 | 02:28:08 | 0.95 | −16.1 | 1.90 | 2.22 | 0.50 | 0.30 | FF | 1.16 |

**Annex B**
Table 2

| 1 | 12 | 13 | 14 | 15 | 16 | 17 | 18 |
|---|---|---|---|---|---|---|---|
| № | $V_A$ [km/s] | $\theta_{Bn}$ [degree] | $V_{sh}$ [km/s] | $C_s$ [km/s] | $V_{fms}$ [km/s] | $M_A$ | $M_{fms}$ |
| 1 | 47.87 | 70.78 | 260.35 | 50.03 | 68.27 | 2.58 | 1.81 |
| 2 | 94.08 | 48.47 | 510.18 | 46.79 | 101.02 | 1.11 | 1.03 |
| 3 | 68.05 | 36.93 | 496.41 | 62.48 | 82.75 | 2.78 | 2.29 |
| 4 | 79.55 | 45.93 | 709.34 | 56.59 | 91.42 | 3.96 | 3.44 |
| 5 | 71.62 | 28.78 | 310.25 | 44.68 | 75.89 | 0.76 | 0.72 |
| 6 | 85.14 | 86.24 | 606.90 | 60.68 | 104.51 | 2.25 | 1.84 |
| 7 | 146.35 | 18.85 | 193.87 | 73.03 | 148.76 | 2.98 | 2.94 |
| 8 | 57.27 | 74.05 | 675.98 | 74.13 | 92.82 | 4.78 | 2.95 |
| 9 | 65.60 | 47.59 | 466.04 | 92.90 | 107.05 | 3.37 | 2.06 |
| 10 | 58.26 | 43.81 | 319.32 | 62.59 | 78.71 | 1.87 | 1.38 |
| 11 | 177.05 | 20.92 | 452.85 | 36.88 | 177.56 | 0.41 | 0.41 |
| 12 | 40.76 | 37.18 | 727.15 | 88.38 | 92.22 | 5.09 | 2.25 |
| 13 | 68.69 | 44.54 | 433.83 | 48.39 | 78.40 | 1.94 | 1.70 |
| 14 | 113.96 | 53.81 | 629.44 | 53.85 | 122.52 | 2.47 | 2.30 |
| 15 | 110.59 | 37.87 | 586.31 | 75.90 | 122.80 | 2.28 | 2.06 |
| 16 | 60.23 | 52.20 | 562.91 | 34.45 | 66.72 | 2.97 | 2.68 |
| 17 | 38.21 | 72.84 | 516.24 | 48.50 | 61.09 | 2.29 | 1.43 |
| 18 | 92.77 | 89.02 | 635.81 | 40.04 | 101.04 | 1.87 | 1.72 |
| 19 | 62.93 | 48.49 | 793.75 | 106.63 | 117.94 | 4.80 | 2.56 |
| 20 | 43.40 | 78.54 | 394.07 | 59.84 | 73.59 | 2.33 | 1.37 |
| 21 | 52.09 | 62.26 | 877.89 | 101.07 | 111.56 | 6.89 | 3.22 |
| 22 | 64.59 | 85.68 | 45.79 | 89.03 | 109.92 | 1.28 | 0.76 |
| 23 | 51.24 | 27.14 | 450.77 | 50.73 | 61.53 | 2.70 | 2.25 |
| 24 | 46.30 | 55.05 | 443.11 | 47.02 | 62.94 | 3.09 | 2.27 |
| 25 | 43.19 | 24.18 | 513.48 | 43.31 | 51.35 | 4.19 | 3.53 |
| 26 | 47.64 | 52.45 | 439.73 | 55.15 | 69.09 | 2.13 | 1.47 |
| 27 | 17.21 | 76.81 | 734.13 | 42.09 | 45.32 | 24.92 | 9.46 |
| 28 | 30.32 | 63.60 | 662.92 | 62.34 | 68.22 | 5.96 | 2.65 |
| 29 | 85.87 | 5.86 | 827.09 | 65.57 | 86.47 | 3.85 | 3.82 |
| 30 | 65.62 | 39.00 | 604.58 | 79.60 | 93.60 | 4.47 | 3.14 |
| 31 | 22.30 | 83.01 | 287.85 | 41.97 | 47.47 | 3.54 | 1.66 |
| 32 | 49.73 | 47.86 | 346.70 | 42.89 | 61.40 | 2.10 | 1.70 |
| 33 | 34.55 | 48.92 | 382.13 | 39.66 | 49.33 | 4.50 | 3.15 |
| 34 | 72.19 | 34.33 | 877.41 | 110.50 | 120.04 | 7.03 | 4.23 |
| 35 | 74.17 | 22.26 | 675.68 | 93.80 | 101.29 | 5.02 | 3.67 |
| 36 | 23.34 | 2.67 | 489.27 | 41.54 | 41.56 | 7.25 | 4.07 |
| 37 | 62.45 | 52.46 | 593.61 | 63.43 | 84.28 | 2.15 | 1.59 |
| 38 | 43.80 | 46.24 | 435.59 | 39.64 | 54.87 | 2.42 | 1.93 |
| 39 | 48.12 | 76.71 | 598.71 | 41.61 | 63.19 | 4.93 | 3.76 |
| 40 | 44.88 | 54.71 | 578.82 | 45.08 | 60.62 | 2.57 | 1.90 |
| 41 | 14.25 | 59.36 | 353.12 | 23.88 | 27.06 | 4.85 | 2.55 |
| 42 | 52.71 | 57.77 | 373.48 | 46.69 | 67.70 | 2.42 | 1.89 |
| 43 | 31.59 | 30.05 | 398.65 | 39.91 | 44.64 | 3.92 | 2.78 |
| 44 | 26.61 | 74.27 | 367.45 | 26.21 | 37.00 | 2.69 | 1.94 |

## Annex C

**Abbreviations in Table 1:**

*Date* indicates the date of registering the shock waves in 2023 (Column 2)
*IP Time* indicates the UTC of registering the shock waves (Column 3)
*Dist* is the distance between the Solar Orbiter and the Sun (Column 4)
*Angle S–E* is the angle between the lines connecting the Solar Orbiter and the Earth with the center of the Sun (Column 5)
$r_B$ is the magnetic-compression factor (Column 6)
$r_N$ is the gas-compression factor (Column 7)
*QF1* and *QF2* are quality factors (Columns 8 and 9)
*IP-Type* indicates the shock-wave type: FF, fast forward; FR, fast reverse; SF, slow forward; SR, slow reverse (Column 10)
$\beta_{us}$ indicates the upstream-zone plasma-beta parameter (Column 11)

**Abbreviations in Table 2:**

$V_A$ is the Alfvenic velocity for the upstream zone (Column 12)
$\theta_{Bn}$ is the angle between the normal line to the shock-wave front and the interplanetary magnetic field vector (Column 13)
$V_{sh}$ is the velocity of the shock-wave front in the rest coordinate system (Column 14)
$C_s$ is the velocity of sound for the interplanetary medium (Column 15)
$V_{fms}$ is the magnetoionic velocity for the upstream zone (Column 16)
$M_A$ is the Alfvenic Mach number for the upstream zone (Column 17)
$M_{fms}$ is the magnetoionic Mach number for the upstream zone (Column 18)